\documentclass[twocolumn]{aastex63}
\usepackage[T1]{fontenc}
\usepackage[normalem]{ulem}
\usepackage[utf8]{inputenc}
\shorttitle{LBTI search for companions and sub-structures in AB Aur}
\shortauthors{Jorquera et al.}

\newcommand{\Lp}{L$^{\prime}$}
\newcommand{\Ks}{K$_{\rm{s}}$}
\newcommand{\Mjup}{M$_{\rm{Jup}~}$}
\newcommand{\Msun}{M$_{\odot}$}

\begin{document}

\title{LBT search for companions and sub-structures in the (pre)transitional disk of AB Aurigae}

\author[0000-0001-6822-7664]{Sebasti\'an Jorquera}
\affiliation{Departamento de Astronom\'ia, Universidad de Chile, Camino El Observatorio 1515, Las Condes, Santiago, Chile}
\affiliation{Universit\'e Grenoble Alpes, CNRS, IPAG, 38000 Grenoble, France}
\author[0000-0001-5579-5339]{Mickaël Bonnefoy} 
\affiliation{Universit\'e Grenoble Alpes, CNRS, IPAG, 38000 Grenoble, France}
\affiliation{Max-Planck-Institut f\"{u}r Astronomie, K\"{o}nigstuhl 17, 69117 Heidelberg, Germany}
\author[0000-0002-8667-6428]{Sarah Betti}
\affiliation{Department of Astronomy, University of Massachusetts, Amherst, MA 01003, USA.}
\author[0000-0003-4022-8598]{Gaël Chauvin}
\affiliation{Unidad Mixta Internacional Franco-Chilena de Astronom\'ia, CNRS/INSU UMI 3386, France}
\affiliation{Universit\'e Grenoble Alpes, CNRS, IPAG, 38000 Grenoble, France}
\author[0000-0003-3306-1486]{Esther Buenzli}
\affiliation{Max-Planck-Institut f\"{u}r Astronomie, K\"{o}nigstuhl 17, 69117 Heidelberg, Germany}
\author[0000-0002-1199-9564]{Laura M. P\'erez}
\affiliation{Departamento de Astronom\'ia, Universidad de Chile, Camino El Observatorio 1515, Las Condes, Santiago, Chile}
\affiliation{N\'ucleo Milenio de Formaci\'on Planetaria (NPF), Chile}
\author[0000-0002-7821-0695]{Katherine B. Follette} 
\affiliation{Physics \& Astronomy Department, Amherst College, 21 Merrill Science Dr., Amherst, MA 01002, USA}
\author[0000-0002-1954-4564]{Philip M. Hinz}
\affiliation{Department of Astronomy and Astrophysics, University of California, Santa Cruz, 1156 High St, Santa Cruz, CA 95064, USA}
\author[0000-0001-9353-2724]{Anthony Boccaletti} 
\affiliation{ LESIA, Observatoire de Paris, Universit\'{e} PSL, CNRS, Sorbonne Universit\'{e}, Univ. Paris Diderot, Sorbonne Paris Cit\'{e}, 5 place Jules Janssen, 92195 Meudon, France}
\author[0000-0002-5407-2806]{Vanessa Bailey}
\affiliation{Jet Propulsion Laboratory, California Institute of Technology, Pasadena, CA 91109, USA}
\author[0000-0003-4614-7035]{Beth Biller}
\affiliation{SUPA, Institute for Astronomy, University of Edinburgh, Blackford Hill, Edinburgh EH9 3HJ, UK} 
\affiliation{Centre for Exoplanet Science, University of Edinburgh, Edinburgh, UK}
\affiliation{Max-Planck-Institut f\"{u}r Astronomie, K\"{o}nigstuhl 17, 69117 Heidelberg, Germany}
\author[0000-0003-3499-2506]{Denis Defr\`{e}re}
\affiliation{Institute of Astronomy, KU Leuven, Celestijnlaan 200D, 3001 Leuven, Belgium}
\author[0000-0002-1031-4199]{Josh Eisner}
\affiliation{Steward Observatory, University of Arizona, 933 North Cherry Ave. Tucson, AZ 85721, USA}
\author[0000-0002-1493-300X]{Thomas Henning}
\affiliation{Max-Planck-Institut f\"{u}r Astronomie, K\"{o}nigstuhl 17, 69117 Heidelberg, Germany}
\author[0000-0002-8227-5467]{Hubert Klahr}
\affiliation{Max-Planck-Institut f\"{u}r Astronomie, K\"{o}nigstuhl 17, 69117 Heidelberg, Germany}
\author[0000-0002-0834-6140]{Jarron Leisenring}
\affiliation{Steward Observatory, University of Arizona, 933 North Cherry Ave. Tucson, AZ 85721, USA}
\author[0000-0003-4475-3605]{Johan Olofsson}
\affiliation{Instituto de Física y Astronomía, Facultad de Ciencias, Universidad de Valparaíso, Av. Gran Bretaña 1111, Playa Ancha, Valparaíso, Chile}
\affiliation{N\'ucleo Milenio de Formaci\'on Planetaria (NPF), Chile}
\affiliation{Max-Planck-Institut f\"{u}r Astronomie, K\"{o}nigstuhl 17, 69117 Heidelberg, Germany}
\author[0000-0001-5347-7062]{Joshua E. Schlieder}
\affiliation{Exoplanets and Stellar Astrophysics Laboratory, Code 667, NASA Goddard Space Flight Center, Greenbelt, MD, 20771, USA}
\affiliation{Max-Planck-Institut f\"{u}r Astronomie, K\"{o}nigstuhl 17, 69117 Heidelberg, Germany}
\author[0000-0001-6098-3924]{Andrew J. Skemer}
\affiliation{Department of Astronomy and Astrophysics, University of California, Santa Cruz, 1156 High St, Santa Cruz, CA 95064, USA}
\author[0000-0001-8671-5901]{Michael F. Skrutskie}
\affiliation{Department of Astronomy, University of Virginia, Charlottesville, VA, 22904, USA}
\author[0000-0002-2190-3108]{Roy Van Boekel}
\affiliation{Max-Planck-Institut f\"{u}r Astronomie, K\"{o}nigstuhl 17, 69117 Heidelberg, Germany}



\begin{abstract}
Multi-wavelengths high-resolution imaging of protoplanetary disks has revealed the presence of multiple, varied substructures in their dust and gas components which might be signposts of young, forming planetary systems. AB Aurigae  bears an emblematic (pre)transitional disk showing spiral structures observed in the inner cavity of the disk in both the sub-millimeter (ALMA; 1.3mm, $^{12}$CO) and near-infrared (SPHERE; 1.5-2.5$\mu$m) wavelengths which have been claimed to arise from dynamical interactions with a massive companion. In this work, we present new deep  \Ks\ (2.16$\mu$m) and  \Lp\ (3.7$\mu$m)  band images of AB Aurigae  obtained with LMIRCam on the Large Binocular Telescope, aimed for the detection of both planetary companions and  extended disk structures. No point source is recovered, in particular at the outer regions of the disk, where a putative candidate ($\rho = 0.681\arcsec, PA = 7.6^{\circ}$) had been previously claimed. The nature of a second innermost planet candidate ($\rho = 0.16\arcsec, PA = 203.9^{\circ}$) can not be investigated by the new data. We are able to derive 5$\sigma$ detection limits in both magnitude and mass for the system, going from 14 \Mjup at 0.3\arcsec (49 au) down to 3-4 \Mjup at 0.6\arcsec (98 au) and beyond, based on the ATMO 2020 evolutionary models. We detect the inner spiral structures  (< 0.5\arcsec) resolved in both CO and polarimetric H-band observations. We also recover the ring structure of the system at larger separation (0.5-0.7\arcsec) showing a clear south-east/north-west asymmetry. This structure, observed for the first time at \Lp~band, remains interior to the dust cavity seen at ALMA, suggesting an efficient dust trapping mechanism at play in the disk. 
\vspace{1cm}
\end{abstract}



\section{Introduction} \label{sec:intro}

The high-resolution capabilities of ALMA  \citep[e.g.,][]{2015ApJ...808L...3A, 2018ApJ...869L..41A, 2020ApJ...889L..24P} in thermal imaging and of ground-based extreme-AO imager instruments in scattered light, e.g. SPHERE \citep{2015A&A...578L...6B, 2016A&A...595A.113S, 2018ApJ...863...44A}, SCExAO \citep{2020ApJ...900..135U}, LBTI \citep{, 2020AJ....159..252W},  have dramatically refined our view of circumstellar disks, revealing substructures in their dust and gas distribution (spirals, concentric annuli, large-scale asymmetries, broken arcs, etc.), and providing direct \citep{2018A&A...617A..44K, 2021arXiv210807123B} or indirect \citep{2019NatAs...3.1109P,2020ApJ...890L...9P} evidence of forming planets embedded in these disks.  

In this context, the Herbig Ae star AB Aurigae ($d = 162.9 \pm 15$ pc, \citeauthor{2018A&A...616A...1G} \citeyear{2018A&A...616A...1G}; $M = 2.4 \pm 0.2$\Msun, \citeauthor{2003ApJ...590..357D} \citeyear{2003ApJ...590..357D}) stands out as one of the most extensively studied young star to date, and a prime example of the complementarity between near-infrared (NIR) and sub-millimeter studies. Polarized light images of the disc in the NIR \citep{2004ApJ...605L..53F,2009ApJ...707L.132P} have revealed multiple spiral structures in the outer regions of the system, extending up to 450 au, together with a warped double ring structure (inner ring radius of 92 au, outer ring radius of 210 au) separated by a gap located at $\approx170$ au \citep{2011ApJ...729L..17H}. Millimeter-wave observations allowed the detection of a large disk in CO, with a central cavity in the dust of 70 au radius \citep{2005A&A...443..945P}. Additionally, \cite{2012A&A...547A..84T}  found counter rotating CO spirals at the outer regions of the disk, with two of them having counterparts in the NIR. 

More recent millimeter and NIR high angular resolution campaigns resolved the innermost regions of AB Aur. Using ALMA, \cite{2017ApJ...840...32T} identified a dust ring with a radius of $\approx120$ au, together with the detection of two clearly defined CO spirals inside the disk cavity, apparently linked to the presence of one or more unseen planetary companions.  Following the ALMA observations, \cite{2020A&A...637L...5B} performed H-band (1.625 $\mu$m) and K-band (2.182 $\mu$m) polarimetric and angular differential imaging (hereafter ADI) observations of AB Aur with SPHERE at the Very Large Telescope (VLT), and reported the detection of two spiral structures inside the disk cavity, together with two apparent point sources at $\approx30$ au and $\approx110$ au. The presence of companions on the system had already been previously discussed based on indirect signatures \citep{2006ApJ...645L..77M, 2008ApJ...679.1574O, 2015ApJ...800...55K}, strongly supporting the results from \cite{2020A&A...637L...5B}. It has also been proposed that the observed inner spirals might be produced due to the interaction with an inclined and eccentric inner binary companion \citep{2020MNRAS.496.2362P}, adding to the complexity of understanding the origin of these structures and  the  ongoing interactions in AB Aur.

We present in two separate studies new deep infrared-imaging (2.2-3.7$\mu$m) observations of AB Aurigae with the Large Binocular Telescope (LBT). The observations benefit from the unique capabilities of the LBT at these wavelengths (low background noise, pyramidal wavefront sensor, real-time redundant observations) and bridge the gap between SPHERE and ALMA observations.  In this study (paper I), we focus on  \Lp\ band observations (3.7$\mu$m) to look for the thermal emission of protoplanets and their surrounding material \citep[circumplanetary-disk and envelope; e.g.,][]{2019MNRAS.487.1248S}. We use both the new \Ks~and \Lp~ band LBTI images to reveal and confirm the faint disk substructures and compare them to those evidenced with SPHERE and ALMA. We present the observations and data processing in Sections \ref{sec:obs} and \ref{sec:red}, respectively. We set new limits on the properties of putative forming planets sculpting the circumstellar dust and gas distribution in Section \ref{sec:detlim}. We investigate disk features in Section \ref{sec:disk} and summarize the findings in Section \ref{sec:conclusions}.

\section{Observations} \label{sec:obs}
\begin{deluxetable*}{cccccccccc}
\label{tab:obs}
\tablecaption{Observing log } 
\tablecolumns{8}
\tablehead{
\colhead{Date} & \colhead{Time (UT)} & \colhead{Target}  & \colhead{Tel. eye} & \colhead{Band} & \colhead{\# exp.}  & \colhead{Used \# exp.}  &  \colhead{$\theta$ ($^{\circ}$)} & \colhead{Seeing (")} & \colhead{Dataset}}
\startdata
2014-02-09 & 02:28 -- 03:58  & AB Aur & SX & \Lp\ & 2700 & 2052 &-69/76 & 0.71 -- 1.06 & 1\\
2014-02-13 & 02:19 -- 03:39  & AB Aur & DX & \Lp\ & 2015 & 1515 &-66/76 & 0.66 -- 1.05 & 2\\
2014-02-13 & 02:19 -- 03:39  & AB Aur & SX & \Lp\ & 2015 & 1558 &\, -66/76$^a$ & 0.66 -- 1.05 & 3\\
2014-02-13 & 04:04 -- 04:35  & HD39925 & DX & \Lp\ & 1210 & 1098 &\,62/75 & 0.73 -- 1.42 & ref2 \\
2014-02-13 & 04:04 -- 04:35  & HD39925 & SX & \Lp\ & 1210 & 1077 &\,62/75  &0.73 -- 1.42 & ref3 \\
2014-02-13 & 04:46 -- 06:53  & AB Aur & DX & \Lp\ & 2500 & 1240 &\,77/70 & 0.68 -- 1.41 & 4\\
2014-02-13 & 04:46 -- 06:53  & AB Aur & SX & \Lp\ & 2500 & 1990 &\,77/70 & 0.68 -- 1.41 & 5\\
2014-02-13 & 07:06 -- 07:31  & HD39925 & DX & \Lp\ & 1020  & 640 &\,73/72 & 1.02 -- 1.83 & ref4 \\
2014-02-13 & 07:06 -- 07:31  & HD39925 & SX & \Lp\ & 1020 & 878 &\,73/72  &1.02 -- 1.83 & ref5 \\
2015-01-04 & 04:08 -- 04:26  & HIP22138 & SX & \Ks\ & 480 & 424 & \,-70/-65  & 0.74 -- 1.03 & ref6 \\
2015-01-04 & 04:42 -- 04:59  & AB Aur & SX & \Ks\ & 480 & 430 & \,-74/-65  & 0.76 -- 1.19 & 6 \\
2015-01-04 & 05:23 -- 05:40  & AB Aur & SX & \Ks\ & 480 & 414 & \,2/60  & unk -- 3.18 & 7 \\
2015-01-04 & 07:13 -- 07:30  & HIP24447 & SX & \Ks\ & 480 & 402 & \,79/80  & 0.68 -- 0.76  & ref7 \\
\enddata
\tablenotetext{a}{AO loop was open for parallactic angles $\theta$ from -61 to 54$^{\circ}$.}
\end{deluxetable*}

Both \Lp\ and \Ks\  band observations of AB Aur (A0Ve, K=4.23, L=3.24) were carried out at the LBT using the LBT Interferometer \citep[LBTI,][]{2008SPIE.7013E..28H} and its adaptive optics (AO) system, over two runs in 2014 and 2015, respectively. Data were taken using the L/M-band InfraRed Camera \citep[LMIRCam,][]{2010SPIE.7735E..3HS}. All observations were performed in pupil-stabilized mode, which allows the use of angular differential imaging \citep[ADI,][]{2006ApJ...641..556M} to subtract the point-spread function (PSF) of the star to reveal its faint surroundings, particularly sharp structures and point-sources. For both bands, PSF reference stars with similar brightness to AB Aur were also observed to apply reference differential imaging (RDI). This strategy can better image any extended circular and spiral structures \citep{2012A&A...545A.111M}, that might be expected for a system like AB\,Aur that is seen close to pole-on. Relevant observing parameters are given in Table \ref{tab:obs}.

\subsection{\Lp\ band observations}
 \Lp\ band ($\lambda_0 = 3.7 \mu\textrm{m}$, $\Delta\lambda = 0.58\mu\textrm{m}$) observations of AB Aur were obtained on the nights of Feb\,8th and Feb\,13th, 2014. On Feb\,8th, observations of AB Aur were taken only with the left (SX) side of the telescope, while both sides were used (in non-interferometric mode) on Feb\,13. 
Individual exposure times were 0.175\,s, five exposures were co-added within the hardware before transferring from the camera. Observations were taken at two nod positions with nodding every 1-2 minutes. On Feb 13th, we also observed HD\,39925 (K5,  K=3.899, L=3.52)  as a PSF reference star, with identical AO settings to allow for referential differential imaging (RDI). We switched between the science target and the reference twice. Because AB Aur transits within 2$^\circ$ of zenith at LBT, the very rapid field rotation caused the AO loop to be unstable for a few minutes. In particular, the left (SX) side AO loop on Feb 13 was completely open between parallactic angles -61 to 54$^\circ$. The right (DX) side obtained continuous observations in that time, but with non-optimal AO correction. Conditions varied from good seeing conditions on Feb 8th, to poor conditions for the final acquisition of the reference star HD\,39925 on Feb, 13th.

For photometric calibration, each of the saturated sequences was immediately followed by unsaturated observations of AB Aur using a neutral density filter with a transmission of 0.9\%.

\subsection{\Ks\ band observations}
\Ks\ band ($\lambda_0 = 2.16 \mu\textrm{m}$, $\Delta\lambda = 0.32\mu\textrm{m}$) observations of AB Aur are described and analized in details in Paper II \citep{2022arXiv220108868B} and used here for completeness. The observations were performed on the night of Jan\,4, 2015, using only the left (SX) side telescope. Individual exposure times were 2s, and observations were taken at two nod positions with nodding every 5 minutes. Two PSF reference stars, HIP\,22138 (G8III, K=4.627, L=4.582) and HIP\,24447 (K0, K=4.219, L=4.136), were observed before and after the observations of AB Aur, respectively, with identical instrument settings.

\section{Data reduction} \label{sec:red}

\subsection{Pre-processing}


We employed a custom-built IDL pipeline \citep{2014A&A...562A.111B} for applying  bad-pixel correction, nod-subtraction, centering of the star by fitting a Moffat profile, aligning the frames, and cropping them to a 300$\times$300 pixels field of view. With a plate scale of 10.7 mas/px \citep{2015A&A...579C...2M} this corresponds to 3.21\arcsec$\times$3.21\arcsec. For AB Aur, frames where the loop was open or where the AO correction was very bad were removed based on a principal component analysis (PCA), with some additional bad frames not detected by this method removed manually. The PCA analysis also revealed that the PSF shape differed significantly between the two nod positions. This was attributed to the dichroic beamsplitter that sends light to both the LMIRCam and the 8-13 $\mu$m camera NOMIC (Nulling Optimized Mid-Infrared Camera)) at LBTI \citep{2016SPIE.9907E..04H} and can cause different internal reflection and diffraction effects at different positions. We therefore treated the data from the two positions separately for the creation and subtraction of the PSF, and recombined them again afterwards. 
The final number of usable frames in each data set is listed in Table \ref{tab:obs}.

\subsection{Subtraction of the stellar halo}

 All the observations were post-processed using the IPAG-ADI pipeline \citep{2012A&A...542A..41C}. We subtracted the PSF for the data which had sufficient field rotation with closed loop through meridian transit (data sets 1 and 2 for AB Aur) using multiple ADI techniques to search for point sources and to reveal asymmetric disk structures. For those data sets where the reference star was observed immediately before or afterwards (data sets 2-7 for AB Aur), RDI processing was applied to reveal the rotationally symmetric disk structures without self-subtraction. 

\subsubsection{Angular differential imaging}\label{ADI}
 We applied on  datasets 1, 2, 6 and 7  classical ADI (cADI), smart ADI (sADI), radial ADI (rADI),  \citep{2012A&A...542A..41C}, Locally Optimized Combination of Images \citep[LOCI, ][]{2007ApJ...660..770L} and Principal Component Analysis \citep[PCA, ][]{2012ApJ...755L..28S} algorithms to remove the stellar halo. The use of these multiple ADI techniques allows for comparison and consistency of the results obtained at different levels of self-subtraction of the disk signal. We refer to \cite{2021AJ....161..146J} for the configuration used for the ADI and LOCI methods. For the PCA method, observations were reduced using three different numbers of modes ($k=1,5,20$), with no radial separation criteria applied for these cases.


For the \Lp\ observations of AB Aur, the parallactic angle coverage was very asymmetric, the duration of the observations was about three times longer at parallactic angle $>+50^\circ$ than at $<-50^\circ$, as can be seen in Figure \ref{fig:abaur_parallactic} (Appendix \ref{App:A}). The rotation between $-50$ and $+50^{\circ}$ was very rapid and only a small range of parallactic angles is covered in the time before and after the rapid rotation. Constructing the PSF as the simple median of the whole cube would result in it being strongly dominated by the majority of frames taken at angles $>50^\circ$. We therefore opted to use the frames at parallactic angles $<0^{\circ}$ as a reference library, to build the PSF for all frames taken at parallactic angles $>0^{\circ}$ and vice versa, using the PCA method. The relatively large average separation in time between the image and its reference PSF in this method results in sub-optimal subtraction. However, it is the only way to avoid heavy self-subtraction at the location of the brightest parts of the disk. The method will essentially subtract out all flux in the image at an angle of $+$ or $-100^\circ$ from the brightest region, but allows us to recover regions with the most asymmetric brightness. 

\subsubsection{Reference differential imaging}
For datasets 2-7 RDI substraction was performed using the PSF reference stars HD39925 (datasets 2-5), HIP 22138 (datasets 6) and HIP 24447 (dataset 7). Observations were reduced using the PCA method with the same configuration as the one described in Sect. \ref{ADI}. For each dataset, observations of each respective reference star were provided as a library, to derive the PSF eigen modes  used for the subtraction of the stellar contribution in the AB Aur science frames.

\begin{figure*}
\centering
\includegraphics[width=18cm]{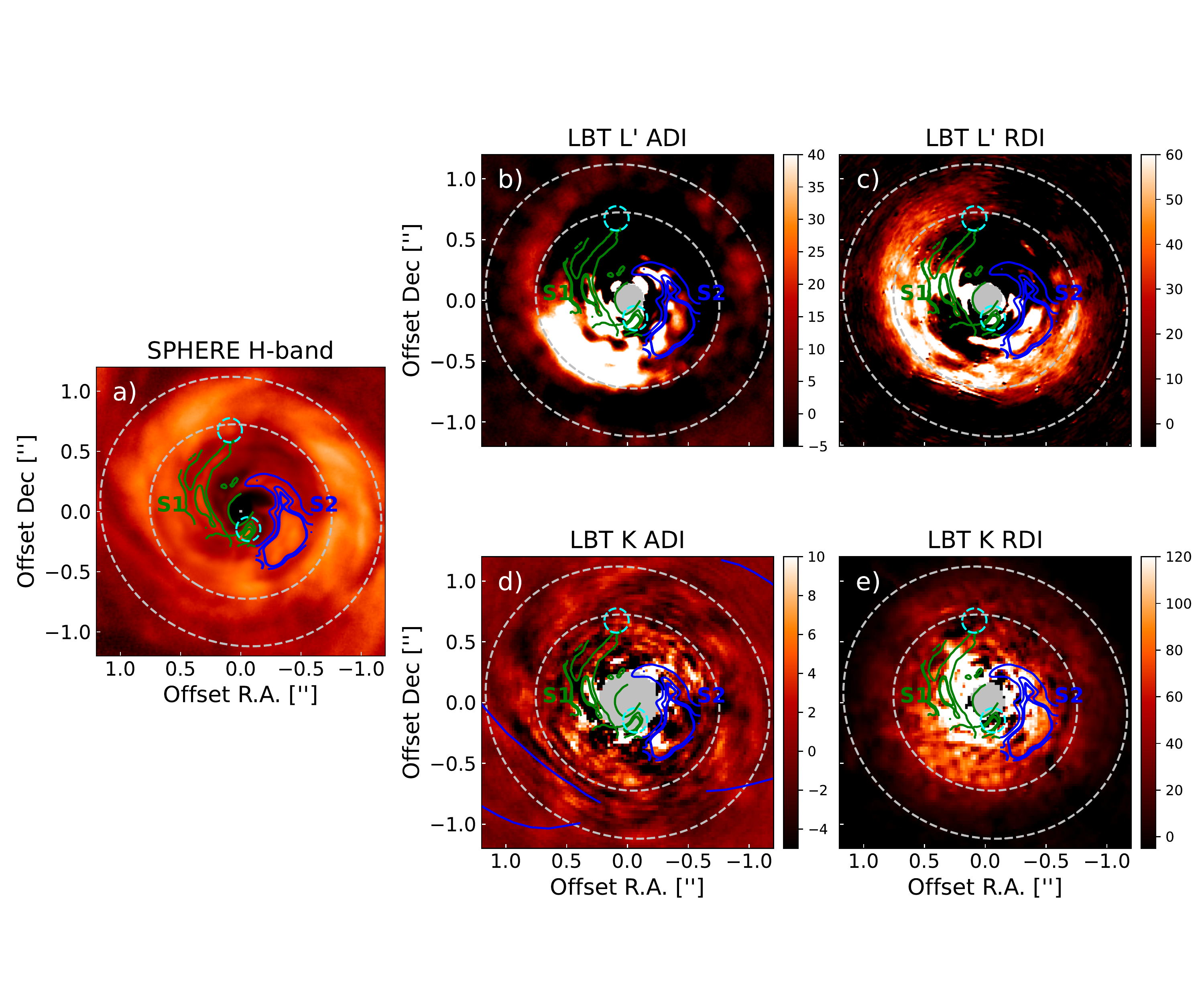}
\caption{Comparison between ADI and RDI reductions for both the \Lp\ (top panels) and \Ks\ band (bottom panels) observations of AB Aur, and the H-band polarized light image obtained with the SPHERE instrument \citep[left panel, ][]{2020A&A...637L...5B}. Images are ordered as follows: a) SPHERE H-band polarimetry image of AB Aur. The polarized intensity has been multiplied by the square of the stellocentric distance ($Q_{phi} \times r^2$) to improve visualization of the structures  b)  Averaged LBT \Lp-band image from dataset 1 and 2 after ADI processing. c) Averaged LBT \Lp-band image from dataset 2 to 5 after RDI processing d) LBT K-band image after ADI processing. e) LBT K-band image after RDI processing. Flux is linear on all the LBT images. Dashed grey lines mark the outer ring reported by \citet{2017ApJ...840...32T}, while the inner spiral arms of the disk reported in \citet{2017ApJ...840...32T, 2020A&A...637L...5B} are denoted by solid green and blue lines. The location of the point sources identified as f1 and f2 by \citet{2020A&A...637L...5B} are marked by a cyan dashed circle. For the case of the K-band ADI reduction, spiral features observed by \citet{2011ApJ...729L..17H} were also marked with solid blue lines in panel (d). Contrast values are given in Analog-Digital Units (ADU) to directly showcase the the results for the different reductions}
\label{fig:comparison}
\end{figure*}

\section{Results}
Figure \ref{fig:comparison} showcases the ADI and RDI reduction of AB Aur for both \Lp\ and K band LBTI observations (top and bottom panels), together with a comparison with the SPHERE H-band polarimetric differential image from \cite{2020A&A...637L...5B} (left panel). No point source appear redundantly in the various LBTI datasets (see Figure \ref{fig:adi_comparison}, Appendix \ref{App:B}) of AB Aur at \Lp~band. Extended emission is however evidenced at \Lp~ in all data-sets for the first time. We present the detection sensitivity inferred from the data as well as a census of the structures seen in the LBTI data below. 

\subsection{Search for planetary companions}
\label{sec:detlim}

\begin{figure*}
\centering
\includegraphics[width=18cm]{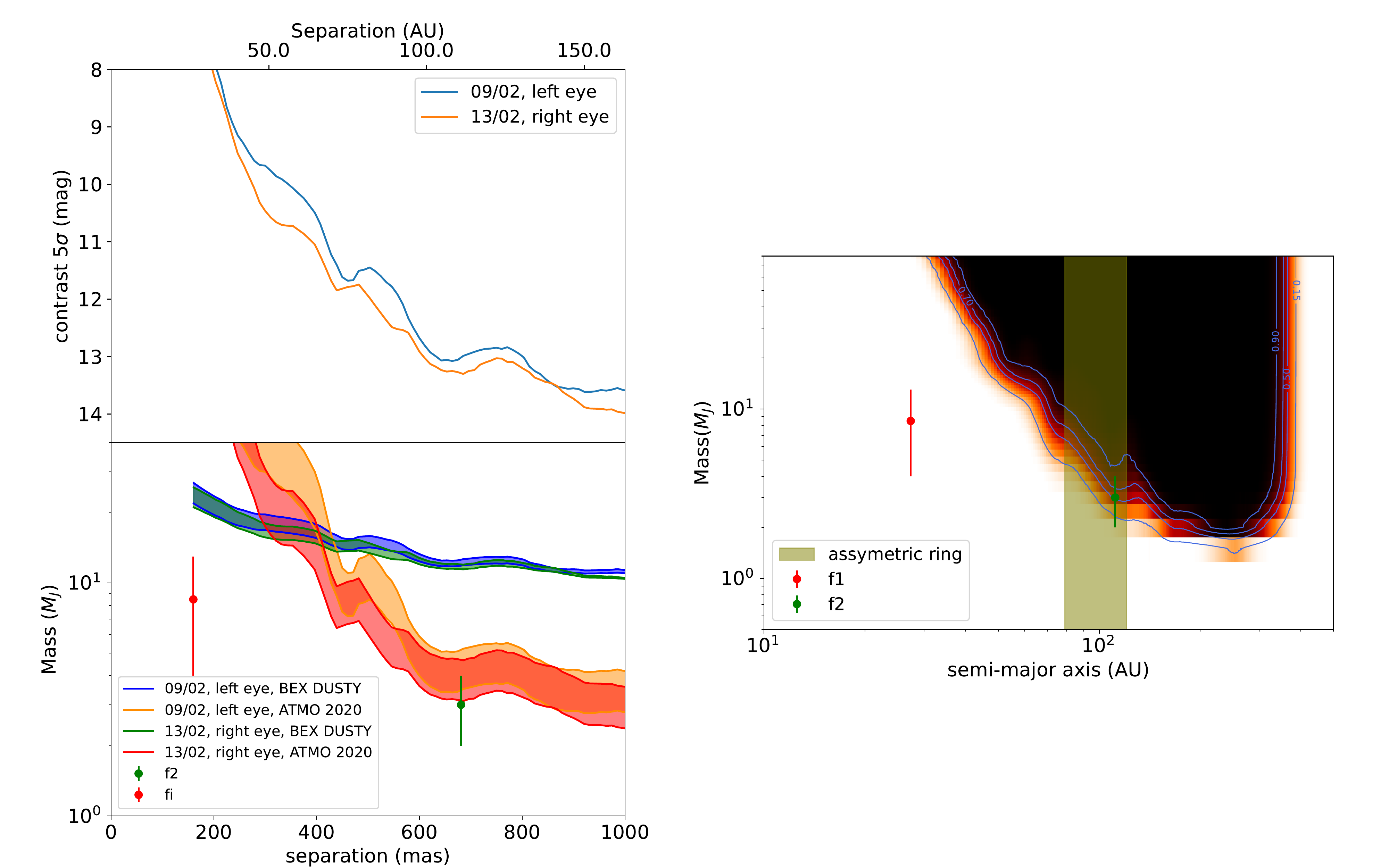}
\caption{Top-left: 5$\sigma$ contrast curve achieved for the observations of AB Aur in \Lp\ taken on two different nights. Bottom-left: Mass limits for point sources for the same set of observations. Shadowed regions correspond to the mass limit variation for an age between 2 Myr (lower curve) and 4 Myr (upper curve)  for AB Aur, based on the ATMOS 2020 and BEX-DUSTY-cold-start evolutionary models. Right: Detection probability map for AB Aur, based on the second night, right eye observations, and an age of 2 Myr for the system. Shaded green region correspond to the location and width of the asymmetric ring. The deprojected semi-major axes for f1 and f2 were derived using the inclination and position angle reported by \cite{2017ApJ...840...32T}. The mass contrast limits are based on COND model predictions. \citep{2003A&A...402..701B}}
\label{fig:abaur_limits}
\end{figure*}

To explore the point-source sensitivity of the LBT observations, pixel-to-pixel noise maps of each ADI reduction were estimated within a sliding box of $1.5\times1.5$\, Full Width at Half-Maximum (\textit{FWHM}) in the processed LBT field of view. We injected regularly spaced fake planets  (every 10 pixels at 3 different position angles, with a flux corresponding to 100\,ADU) in the original data cubes  to evaluate the flux losses caused by the ADI process \citep[e.g.,][]{2010A&A...509A..52C, 2013A&A...553A..60R}. The final $5\sigma$ contrast maps were obtained using the pixel-to-pixel noise maps divided by the flux loss and normalized by the relative calibration with the primary star, and were also corrected from small number statistics following the prescription of \citet{2014ApJ...792...97M} to adapt our $5\sigma$ confidence level at small angles. 

The detection limits for datasets 1 and 2, obtained from the cADI processing, are shown in Figure \ref{fig:abaur_limits}. On the second night, the achieved $5\sigma$ mass limits were up to 1 mag better than on the first night, depending on the separation. We reached a contrast of 10 mag at 0.25\arcsec, 12.5 mag at 0.5\arcsec, and 14 mag at 1\arcsec. The 5$\sigma$ contrast curves were then translated to mass limits based on the mass-luminosity relation for giant planets predicted by the ATMO 2020 \citep{2020A&A...637A..38P} and BEX-DUSTY-cold-start \citep{2017ApJ...836..221M, 2021A&A...652A.101A} evolutionary models. Mass limits were derived considering either a 2 and 4 Myr system. For the more optimistic scenario of 2 Myr of age, the mass limits for companions around AB Aur from the ATMO 2020 models translate to: 3-4 \Mjup beyond 0.6\arcsec ($\sim 100$ au), 6 \Mjup at 0.5\arcsec (81 au), and 14 \Mjup at 0.3\arcsec (50 au). The inner working angle of the observations is approximately 0.2\arcsec (28 au), where we would have been sensitive to brown dwarf companions with masses above $\approx$35 \Mjup. In the case of the BEX-DUSTY-cold-start models we recover mass limit of 25-20 \Mjup between 0.2\arcsec and 0.4\arcsec, 20-15 \Mjup between 0.4\arcsec and 0.6\arcsec and 15-11 \Mjup beyond 0.6\arcsec. Although this results clearly indicate that we are only sensitive the high mass objects arising from this cold-start or low entropy scenario, it is important to point out that recent works \citep[e.g.,][]{2017ApJ...836..221M, 2017ApJ...834..149B} have showed that cold start is actually unlikely to occur for giant planet formation and that these objects are more likely associated with a high initial entropy \citep{2017A&A...608A..72M, 2019ApJ...881..144M}. Because of this, further analysis regarding the possible presence of planetary companions on the system is based on our derived ATMO 2020 detection limits exclusively.

Based on our obtained mass limits, we derive a detection probability map for the second night, right eye observations as they provide a higher sensitivity to point sources, and assuming an age of 2 Myr. This map was obtained by using the Multi-purpose Exoplanet Simulation System (MESS) code, a Monte Carlo tool for the predictions of exoplanet search results \citep{2012A&A...537A..67B}. We generated a uniform grid of masses and semi-major axis in the interval [0.5, 80] \Mjup and [10, 1000] AU with a sampling of 0.5 \Mjup and 1 AU respectively. For each point in the grid, $10^4$ orbits were generated with a fixed inclination and position angle \citep[based on the inclination and position angle of the disk from ][]{2017ApJ...840...32T}, but randomly oriented in space from uniform distributions in  $\omega, e < 0.1$ and $M$, which correspond to the argument of periastron with respect to the line of nodes, eccentricity, and mean anomaly, respectively. The detection probability map is then built by counting the number of detected planets over the number of generated ones, by comparing the on-sky projected position (separation and position angle) of each synthetic planet with the 2D mass maps at $5\sigma$.\\

 \citet{2017ApJ...840...32T} proposed that the observed inner CO spirals might be triggered by the presence of a planetary companion at either 60-80 au or 30 au from the central star, while \cite{2020A&A...637L...5B} recovered the signal of two apparent point sources located at a separation of 0.16\arcsec ($\sim26$ au; f1) and 0.681\arcsec ($\sim111$ au; f2), with f1 seemingly associated with one of the observed spirals in H-band, consistent with the predictions from the ALMA CO observations. The location and estimated masses of f1 and f2 are reported in Figures \ref{fig:comparison}  and  \ref{fig:abaur_limits}. None of these point sources are recovered in any of our ADI and RDI reductions at \Lp\ and \Ks\ bands. As we can only resolve objects down to a separation of approximately 0.2\arcsec, detection of f1 is not possible on our \Lp\ and \Ks\ observations, which also applies to the possible binary companion proposed by \citet{2020MNRAS.496.2362P}. In the case of f2, the constrast derived by \cite{2020A&A...637L...5B} is consistent with a mass ranging from 2 to 4 \Mjup based on the AMES-COND models. \cite{2020A&A...637L...5B} also obtained a preliminary mass estimation of $\sim 3$ \Mjup based on dynamical considerations derived by \cite{1980AJ.....85.1122W}, which provides a relation between location of the inner edge of the observed cavity, the star mass, and the planet's mass and distance. 
 
\begin{figure}
\epsscale{0.9}
\plotone{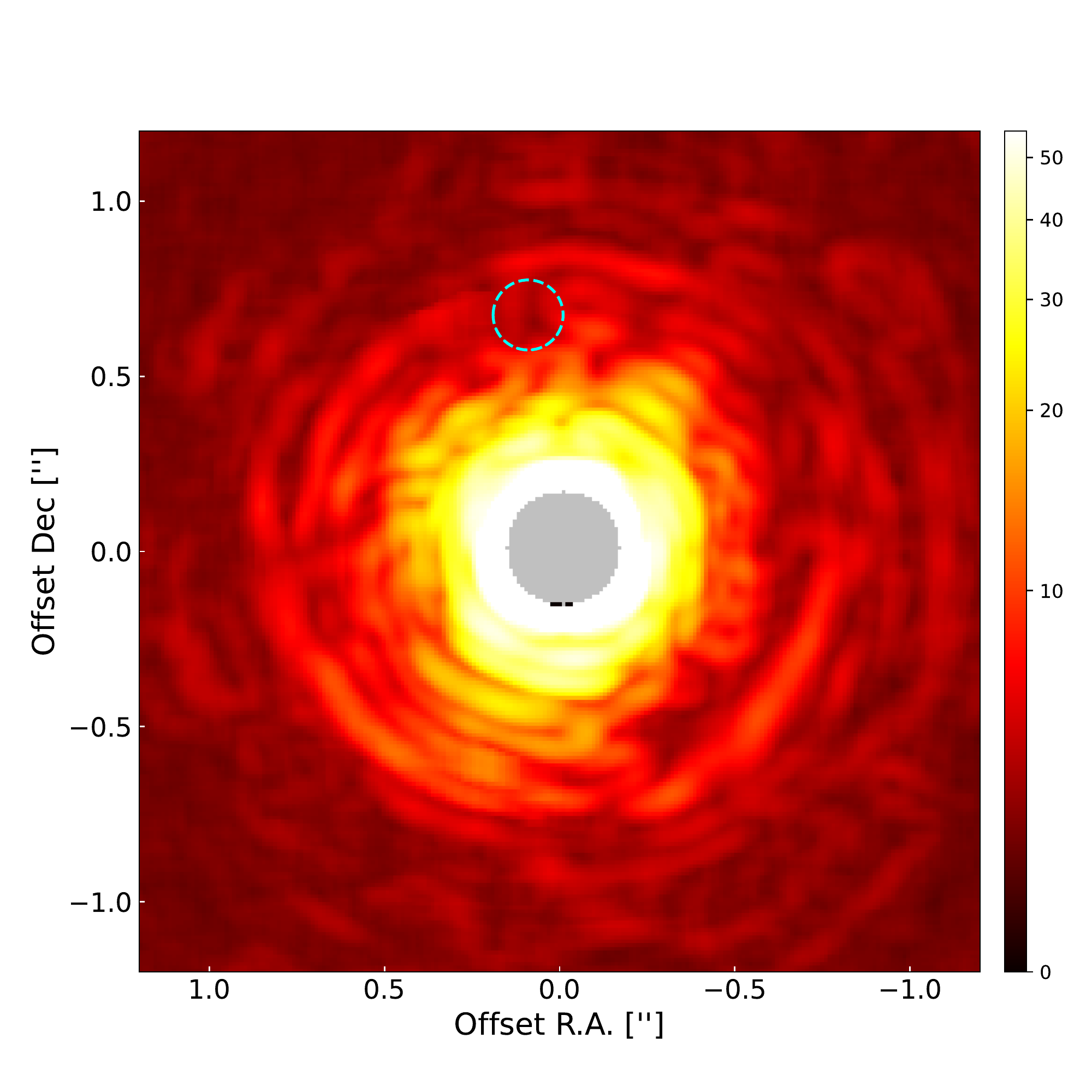}
\caption{Mass detection map obtained for the second night of observations of AB Aur. The color bar indicates the minimum detectable mass  in \Mjup. The cyan circle corresponds to the location of f2.}  
\label{fig:abaur_psfsub}
\end{figure}

 This predicted mass lies in the 5 $\sigma$ mass limit range derived from the second night of LBTI \Lp~observations and has a detection probability of $50\%$ at the deprojected position of f2, for an age of 2 Myr. At first glance, this detection probability is not enough to properly confirm or discard the existence of f2. However, it is important to take into account that the derived detection probability for a given semi-major axis (orbit) does not imply that the detection probability is the same at every point of that orbit. This is because this probability is only based on the count of detected planets along the full extension of the orbit, not considering where in the orbit they were specifically recovered. We can solve this problem by looking at the bi-dimensional mass detection limit at the specific location of f2 (Fig. \ref{fig:abaur_psfsub}). At this location, we obtained a 5$\sigma$ mass detection limit of $\sim 3.16$ \Mjup, very close to the mass limit from \cite{2020A&A...637L...5B} obtained at shorter wavelengths (hence, likely more impacted by disk foreground extinction) and implying that the detection probability at this location should be higher than $50\%$. Based on these results, the lack of detection of f2 suggests it is not a real companion.

\subsection{Disk geometry and sub-structures}
\label{sec:disk}

We detect disk emission at \Lp~band up to $\sim$1" separation. The innermost extended emission (< 0.5 \arcsec) at the east and west sides of the disk are in good agreement with the location of the two main spirals detected by \cite{2020A&A...637L...5B}, with the western spiral (S2) being more clearly detected, while only the starting part of the eastern spiral (S1) appears to be recovered. The consistency of the appearance of these structures for both reduction methods (ADI and RDI), as well as for each of the independent ADI reductions (see Figure \ref{fig:adi_comparison}, Appendix \ref{App:B}) suggests that the recovered signal effectively corresponds to the inner spirals while the fragmented appearance of S2 is most likely caused by non-optimal halo subtraction. 

A brighter ring-like emission centered at a radius of $\approx 0.56 \arcsec$ is evidenced in the \Lp~ and \Ks~ LBTI images. It corresponds to a similar structure evidenced in the Subaru/HiCIAO  \citep{2011ApJ...729L..17H} and  VLT/SPHERE polarized intensity H-band observations of \cite{2020A&A...637L...5B}  and shows a strong brightness asymmetry. This asymmetry is seen in the HiCIAO images but not recovered in the SPHERE observations. This suggests it to be caused by finer substructures left unresolved in our observations and \cite{2011ApJ...729L..17H} ones.


The ADI and RDI post-processed \Ks-band images correspond to the d) and e) panels of Fig. \ref{fig:comparison}, respectively. The inner regions ($<$0.5") in the ADI images appears to be polluted by speckles. Instead, this reduction appears to be more sensitive at the outer parts of the disk, as faint emission from different structures, and particularly spirals, appears to be recovered, which is in good agreement with previously observed spirals \citep{2011ApJ...729L..17H}. Finally, the RDI reduction reveals the ring-like emission evidenced at H and \Lp~bands. The structure also shows the asymmetry reported at \Lp-band, hence possibly caused by the coarser angular resolution of these observations.  

The ring-like structure lies at H, \Ks, and \Lp~bands within the dust ring detected in the ALMA continuum observations (1.3 mm) \citep{2017ApJ...840...32T} as shown in Fig. \ref{fig:comparison}. This difference between observed structures at different wavelengths can be explained by considering the effect of dust trapping in a disk, that can be triggered by planet-disk interactions of one or more planetary companions. The presence of a massive planet, such as the case of f1, induces a pressure enhancement on the gas distribution of the disk, which traps large particles in this pressure maxima, while smaller particles are allowed to drift into the inner cavity that is depleted of large grains \citep{2013A&A...560A.111D, 2015A&A...573A...9P, 2018A&A...617A..44K}. 
This could explain why a sharp inner edge can be observed for the dust ring at 1.3 mm, that traces large particles, while dust grains of smaller size can still be detected in the cavity, producing the features observed in the near-infrared. 

Another possible explanation for the differences in wavelength of the observed structures, specifically the asymmetric ring-like structure, arises when considering the time variability of Herbig Ae/Be stars. It has been showed that these types of systems present irregular photometric \citep{2002A&A...384.1038E} and spectroscopic \citep{2011A&A...529A..34M} variations, which has been attributed to shadowing from the inner ring structure \citep{2018A&A...620A.128V, 2019ApJ...887L..32C}. If that is the case for AB Aur, then it is possible that the observed differences are only product of this variabilty between the different epochs of observation.

Although out of the scope of this study, further modelling of the disk would allow to determine the degree asymmetry of the phase function for the dust scattering \citep[][]{1941ApJ....93...70H} and subsequently the dust grain size of the observed region,  and to put strong constraints on the properties of both possible planetary companions and the protoplanetary disk, as dust trapping is heavily dependant on those parameters. At the same time, further observations at different NIR bands at similar epochs would allow to properly determine the possible time variability of AB Aur and further identify if the differences between observed structures are also influenced by this variations. 


\section{Conclusions}
\label{sec:conclusions}
In this work we present \Lp\ and \Ks\ band observations of AB Aur obtained with LBT and processed using multiple flavours of ADI and RDI methods, aimed for the detection of possible planetary companions and characterization of the disk structures, with the intention to bridge the results obtained from SPHERE NIR and ALMA sub-millimeter observations of the same target. Our main results can be summarized as follows.

\begin{itemize}


\item No planetary companions were detected in our observations. Contrast ($5\sigma$ magnitude) and mass detection limits were derived from the \Lp\ observations of the target, serving as a first estimator for the detectability of planetary companions in the system. Our results translate to 3-4 \Mjup beyond 0.6\arcsec up to 14 \Mjup at 0.3\arcsec (50 au). Comparison with previous estimations on the location of possible companions at 0.16\arcsec (26 au; f1) and 0.681\arcsec (111 au; f2) reveal that f1 can not be resolved in our observations, based on the degradation of the ADI and RDI methods at this separation. In the case of f2 we set an upper mass limit of 3.16 \Mjup for a non-extincted massive planet based on our \Lp\ detection limits.

\item Part of the spiral structures resolved in both CO and polarimetric H-band observations at the inner region (< 0.5 \arcsec) of the disk were recovered from the \Lp\ observations, although poorly resolved due to the performance of the ADI and RDI methods at this separations and wavelengths with LBT. 

\item We report the first detection of the ring structure of the system at \Lp~ which shows a brightness assymetry also noticed in HiCIAO H and the LBTI \Ks-band observations, but not in the SPHERE H-band data. This apparent inconsistency might be cause by the different angular resolution of the various ground-based observations. 

\item We discuss dust trapping in the disk and time variability of Herbig Ae/Be stars as the possible explanation for the difference in asymmetry observed between the different bands. Further observations and modelling are needed to properly determine the effect that each of these phenomena have in th observed structures.

\item The ADI reduction \Ks\ band observations also allows to resolve the outer structures of the disk, revealing the presence of spiral arms which are in very good agreement with previously observed spiral structures.%

\end{itemize}

The \Lp-band images of AB Aur add to the limited sample of disks resolved at these wavelengths from the ground \citep[e.g., ][]{2019ApJ...882...20W, 2019ApJ...877L...3C, 2020AJ....159..263W, 2020AJ....159..252W}. Such challenging observations should increase steadily with the recent \Lp-band imagers operating on 8-m class telescopes fed by adaptive-optics systems, better adapted to the observations of red dust-enshrouded host stars (LBT/LMIRCam, Keck/NIRC2, VLT/ERIS-NIX). 
 
\acknowledgements
S.J.\ acknowledges support from the National Agency for Research and Development (ANID), Scholarship Program, Doctorado Becas Nacionales/2020 - 21212356.
EB was supported by the Swiss National Science Foundation (SNSF). 
L.P. gratefully acknowledges support by the ANID BASAL projects ACE210002 and FB210003, and by ANID, -- Millennium Science Initiative Program -- NCN19\_171.
J.\,O. acknowledges support by ANID, -- Millennium Science Initiative Program -- NCN19\_171, from the Universidad de Valpara\'iso, and from Fondecyt (grant 1180395)
This research has made use of the SIMBAD database, operated at CDS, Strasbourg, France, and of NASA's Astrophysics Data System Bibliographic Services. This work benefited from the support of the project FRAME ANR-20-CE31-0012 of the French National Research Agency (ANR).

\appendix

\section{Parallactic angle variation}\label{App:A}
We present the parallactic angle coverage for the 7 independent sets of observations of AB Aur. 

\begin{figure}
\epsscale{0.9}
\plotone{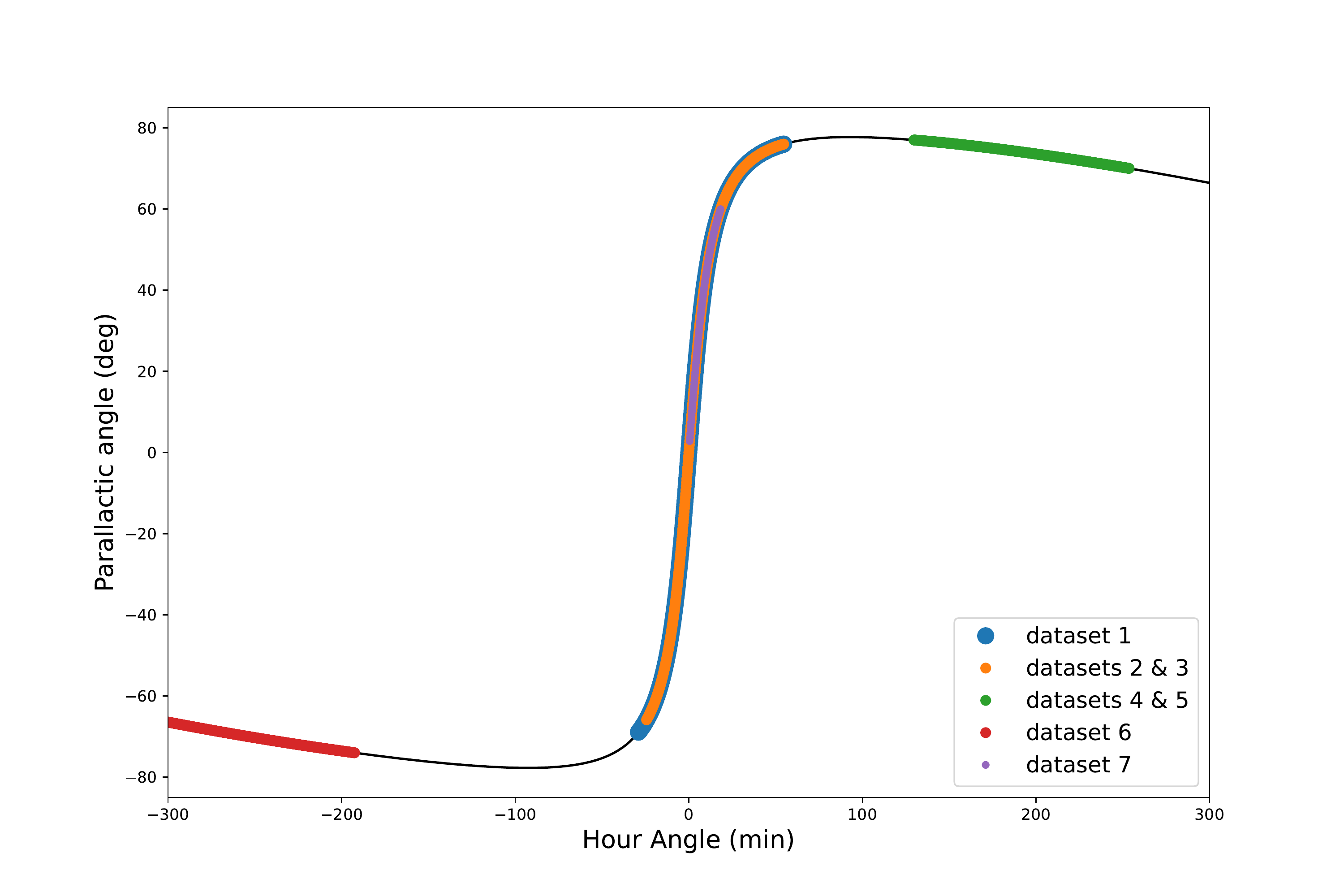}
\caption{Parallactic angle variation of the 7 independent datasets of observations of AB Aur. The black line indicates the expected variation of the parallactic of the object during a normal night, taking into account the rotation of the FOV due to the use of the pupil-stabilized mode.} 
\label{fig:abaur_parallactic}
\end{figure}

\newpage

\newpage
\section{ADI reduction}\label{App:B}

\begin{figure}
\epsscale{0.9}
\plotone{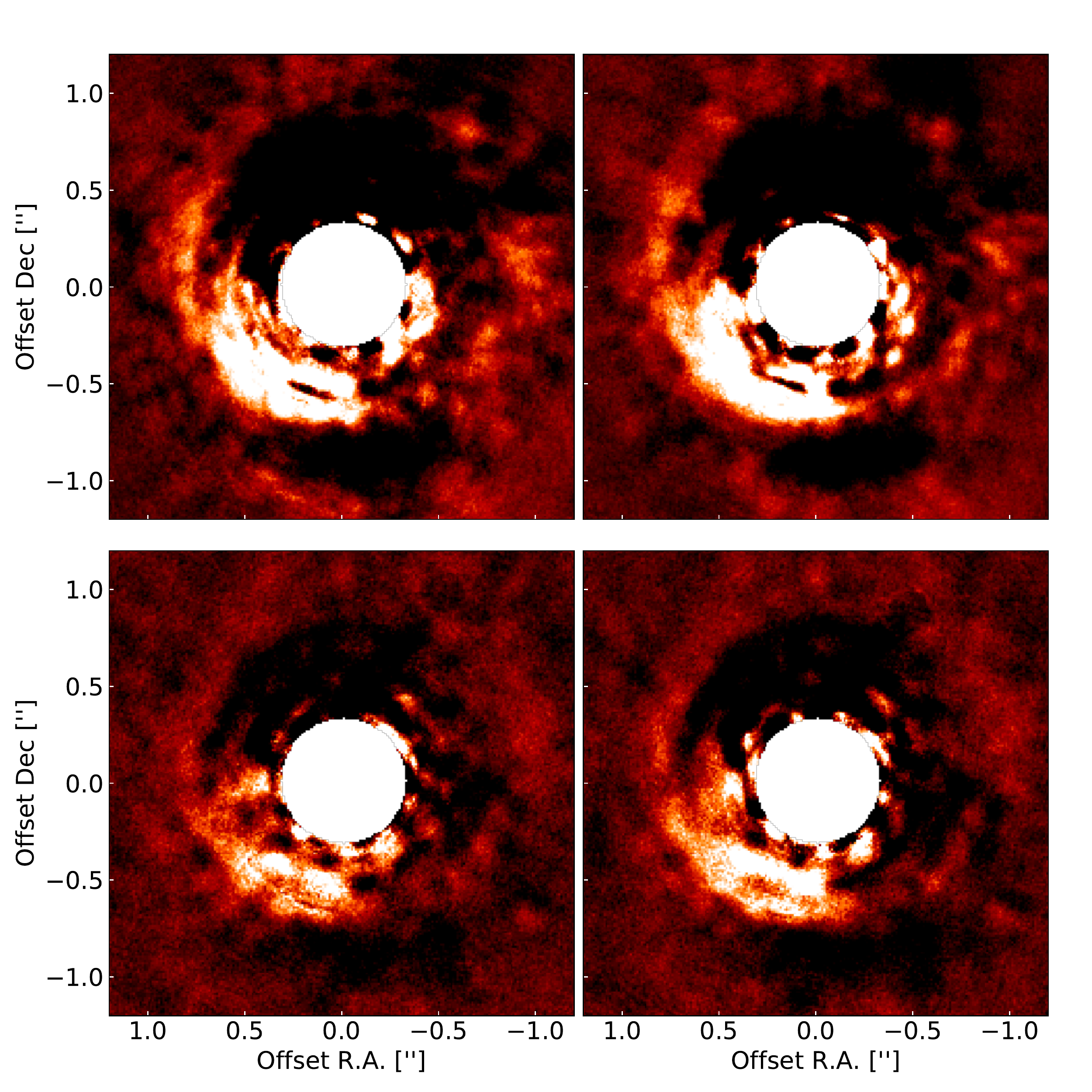}
\caption{Top: ADI reduction from ABAur observations obtained using the left side of the telescope. Bottom: ADI reduction from AB Aur observations obtained using the right side of the telescope. Left (Right) images were constructed using observations before (after) the meridian passage.}  
\label{fig:adi_comparison}
\end{figure}

Figure \ref{fig:adi_comparison} showcases the ADI reduction for datasets 1 and 2, with the observations before and after the meridian passage, reduced separately for each case. This allows to check for consistency on the recovered structures and to discard possible artifacts. In all cases, we recover part of the inner spirals S1 and S2, as well as the bright part of the disk on the southeast region.
\newpage

\newpage

\bibliography{imaging}{}
\bibliographystyle{aasjournal}

\end{document}